\documentclass[12pt]{article}
\usepackage[dvips]{graphicx}

\usepackage{epsfig,amsmath,amssymb,verbatim,mathrsfs}

\def\beq{\begin{eqnarray}}
\def\eeq{\end{eqnarray}}
\def\bea{\begin{eqnarray}}
\def\eea{\end{eqnarray}}
\def\bp{\beta_+}
\def\bm{\beta_-}
\def\bl{\bar{\lambda}}
\def\tl{\tilde{\lambda}}
\def\la{\lambda_a}
\def\lb{\lambda_b}
\def\lp{\lambda_+}
\def\lm{\lambda_-}

\begin{document}

\setlength{\baselineskip}{0.2in}

\begin{titlepage}
\noindent

\vspace{10cm}
\flushright{August 2010}
\vspace{1cm}
\begin{center}
  \begin{Large}
    \begin{bf}
Broken Symmetry as a Stabilizing Remnant 
     \end{bf}
  \end{Large}
\end{center}
\vspace{0.2cm}

\begin{center}

\begin{large}
Sandy S. C. Law$^{(a)}$ and Kristian L. McDonald$^{(b),(c)}$\\
\end{large}
\vspace{1cm}
  \begin{it}
$(a)$ Department of Physics, Chung Yuan Christian University, \\
Chung-Li, Taiwan 320, Republic of China.\\Email: slaw@cycu.edu.tw \vspace{0.5cm}\\
$(b)$ TRIUMF,
4004 Wesbrook Mall, Vancouver, BC V6T 2A3, Canada.\\\vspace{0.5cm}
$(c)$ Max-Planck-Institut f\"ur Kernphysik,\\
 Postfach 10 39 80, 69029 Heidelberg, Germany.\\
Email: Kristian.McDonald@mpi-hd.mpg.de
\end{it}
\vspace{0.5cm}

\end{center}

\begin{abstract}
 The Goldberger-Wise mechanism enables one to stabilize the length of
 the warped extra dimension employed in Randall-Sundrum models. In
 this work we generalize this mechanism to models with multiple warped
 throats sharing a common ultraviolet brane. For independent throats
 this generalization is straight forward. If the throats possess a
 discrete interchange symmetry like $Z_n$ the stabilizing dynamics may
 respect the symmetry, resulting in equal throat lengths, or they may
 break it. In the latter case the ground state of an initially
 symmetric configuration is a stabilized asymmetric configuration in
 which the throat lengths differ. We focus on two- (three-)  throat
 setups with a $Z_2$ ($Z_3$) interchange symmetry and present
 stabilization dynamics suitable for either breaking or maintaining
 the symmetry. Though admitting more general application, our results
 are relevant for existing models in the literature, including the two
 throat model with Kaluza-Klein parity and the three throat model of
 flavor based on a broken $Z_3$ symmetry.
\end{abstract}

\vspace{1cm}

\end{titlepage}
%\pacs{PACS numbers: }
%]

%\setcounter{footnote}{1}
\setcounter{page}{1}
%\setcounter{figure}{0}
%\setcounter{table}{0}

%\tableofcontents

\vfill\eject

%\end{comment}

%%%%%%%%%%%%%%%%%%%%%%%%%%%%%%%%%%%%%%%%%%%%%%%%%%%%%%%%%%%%%%%%%%%%%
\section{Introduction\label{sec:intro}}
The ultraviolet (UV) sensitivity of the Higgs mass in the standard model (SM) makes it difficult to understand how the Higgs can remain light in the presence of heavy new physics. Mass corrections resulting from top quark loops can dominate the Higgs mass, dragging it up to the cutoff scale and thereby necessitating a fine tuning to preserve a light Higgs. This ``hierarchy problem'' provides perhaps our best indication that new physics is likely to appear at the TeV scale and it is hoped that the LHC will soon shed light on this matter. 

The Randall-Sundrum (RS) model~\cite{Randall:1999ee} provides a
candidate solution to the hierarchy problem. This  framework allows
one to generate natural scale hierarchies as the infrared (IR) scale
is realized as a warped down incarnation of the Planck scale, $M_{IR}\sim e^{-k L}M_{Pl}$, where $k$ ($L$) is the curvature (length) of the warped extra dimension. Provided one can naturally realize the hierarchy $kL\simeq \mathcal{O}(10)$ the weak scale can be generated with $M_{IR}\sim$~TeV and the Higgs mass protected from large corrections. The relationship between the curvature and $L$ is determined by the dynamics that stabilize the extra dimension and the question of whether the RS model naturally realizes the weak/Planck hierarchy translates into the need for stabilization dynamics that naturally generate $kL\simeq \mathcal{O}(10)$.

Goldberger and Wise (GW)~\cite{Goldberger:1999uk} have presented a mechanism that successfully generates $kL\simeq \mathcal{O}(10)$ without the need for input parameter hierarchies, thereby showing that the RS model provides a genuine solution to the hierarchy problem. The GW mechanism employs a bulk scalar field with brane localized potentials that force the scalar to acquire distinct nonzero values at the branes. The resulting interplay between the shearing energy (which prefers the extra dimension to be large) and the potential energy (which tends to shrink the extra dimension) of the background scalar solution stabilizes the extra dimension  at a fixed finite value. Alternative methods for stabilizing $L$  have also appeared~\cite{Garriga:2002vf}.

Besides the Planck and weak scales there may be other scales present in nature. Examples of such scales occur generically in models with gauge UV completions of the type $\mathcal{G}_{UV}\supset\mathcal{G}_{SM}$, where $\mathcal{G}_{UV}$ may be a grand unified gauge group, or some other gauge extension of the SM, that is broken to $\mathcal{G}_{SM}$ at a posited high energy scale. The flavor sector of the SM is also suggestive of new scales in nature if the Yukawa couplings emerge as powers of a dimensionless ratio $\langle \phi_f\rangle/\Lambda$ for some flavon fields $\phi_f$ and a cutoff $\Lambda$. Further scales may exist if the dark or hidden sector of the universe is not directly connected to the weak scale. The dark matter itself may acquire its mass by a distinct means to the SM fields, but even if the dark matter obtains a weak scale mass it can couple to forces whose mass scale is much lighter~\cite{ArkaniHamed:2008qn}.

 If the RS approach to the weak/Planck hierarchy is realized in
 nature it is natural to ask if additional scales can also be
 accommodated in this framework. The gravitational background of the
 RS model  can be referred to as a warped throat and one can extend
 the RS gravitational background by considering multiple warped
 throats glued together at a common UV
 brane~\cite{Cacciapaglia:2006tg}. If the IR scales of the distinct
 throats differ the warping that realizes the weak/Planck hierarchy in
 RS models can also be employed to obtain additional mass
 scales.\footnote{See~\cite{Gherghetta:2010cq} for an alternative way
   to realize a sub-TeV hidden sector scale in addition to the
   weak/Planck hierarchy in the RS framework.} Such setups may have a
 number of applications, including
 electroweak~\cite{Cacciapaglia:2005pa},
 flavor~\cite{Abel:2010kw,Plentinger:2008nv},
 leptogenesis~\cite{Bechinger:2009qk}, axion~\cite{Flacke:2006ad} and
 hidden sector physics~\cite{Gripaios:2006dc,McDonald:2010iq}. If the
 IR scales of multiple throats are related by symmetry one can also
 motivate dark matter candidates~\cite{Agashe:2007jb}. As in RS models
 one must ensure that the throat lengths are suitably stabilized to
 naturally generate multiple hierarchical scales in multithroat models. 

In this work we generalize the GW mechanism to models with multiple
warped throats sharing a common ultraviolet brane. We consider both
independent throats\footnote{In this work an ``independent throat''
  refers to a throat that is part of a multithroat background but is
  not related to any of the other throats by an interchange symmetry.}
and throats possessing a discrete interchange symmetry like $Z_n$. In
the former case the generalization is straight forward. In the latter
case the stabilizing dynamics may respect the symmetry, resulting in
equal throat lengths, or they may break it, producing an asymmetric
configuration in which the throat lengths differ. We shall focus on
two- (three-)
throat setups with a $Z_2$ ($Z_3$) interchange symmetry
and present stabilization dynamics suitable for either breaking or
maintaining the symmetry. Though admitting more general application,
our results are relevant for existing models in the literature and we
shall draw attention to some of these as appropriate. 

One of the main
points that we seek to bring to the readers attention is
the new approach to discrete symmetry breaking afforded by our
constructs. Models with discrete
symmetries in four dimensions typically break the symmetry spontaneously
with a weakly coupled Higgs or via
strong dynamics at some energy scale. This symmetry
breaking scale usually maps to some IR scale in the theory,
like the top mass or some other fermion mass scale in models with discrete
flavour symmetries, but must also be sufficiently shielded from SM
fields to ensure compatibility with observations. One interesting aspect of discrete symmetry breaking
via GW scalars is that the symmetry breaking can occur entirely at a
high energy scale, like the Planck
scale, through UV localized dynamics and subsequently feed into the IR through
the emergence of distinct IR scales. This is interesting in the case
of a discrete flavour symmetry, like that recently discussed
in~\cite{Abel:2010kw}, as the SM fields need not couple
directly to the source of symmetry breaking in the UV. They may instead
couple to IR flavour fields in a flavour symmetric way and yet exist
as part of a flavour asymmetric theory in the IR. We suspect that our
ideas may admit interesting applications to flavour model building and
in particular geometric throat arrangements may provide an alternative
extra dimensional approach to discrete flavour symmetries to that
presented in~\cite{Altarelli:2006kg}. 

Before proceeding we note that the field-theoretic approach to
multithroat models~\cite{Cacciapaglia:2006tg} can be motivated by the fact that string realizations of the RS model can contain additional warped throats~\cite{StringThroats}. In the string picture these emerge from the compact space that acts as the UV brane in the RS approach and the notion of multiple throats glued together in the UV serves to model this string picture. Earlier phenomenological applications of the multithroat setup can be found in~\cite{Dimopoulos:2001ui}.

The organization of this paper is as follows. In Sec.~\ref{sec:GW_example} we present a single throat calculation to remind the reader of the GW methodology and set our notations. We generalize the GW mechanism to two independent throats in Sec.~\ref{sec:2ind_throats}. We consider two throats related by a $Z_2$ symmetry in Sec.~\ref{sec:GW_z2} and three throats related by a $Z_3$ in Sec~\ref{sec:GW_z3}. In both cases we present symmetry preserving and symmetry breaking GW mechanisms. Sec.~\ref{sec:more_than_three} contains some comments on models with $n>3$ throats and the paper concludes in Sec.~\ref{sec:conc}.
%%%%%%%%%%%%%%%%%%%%%%%%%%%%%%%%%%%%%%%%%%%%%%%%%%%%%%%%%%%%%%%%%%%%
\section{GW Mechanism in a Single Throat\label{sec:GW_example}}
We would like to present an example calculation to demonstrate the approach of GW. Rather than summarizing the single throat calculation whose details can be found in~\cite{Goldberger:1999uk} we present a slightly modified calculation which, despite failing to successfully stabilize the length of the extra dimension, serves to demonstrate the methodology  and sets our notations. It is also useful in helping us understand features that emerge in some of the multithroat calculations that follow. 

The metric we employ is defined by the interval
\bea
ds^2=e^{-2ky}\eta_{\mu\nu}dx^\mu dx^\nu -dy^2\equiv G_{MN}dx^M dx^N,
\eea
where  $M,N,..$ ($\mu,\nu,..$) are the 5D (4D) Lorentz indices, the extra dimension is labeled by $y\in[0,L]$ and the UV (IR) brane is located at $y=0$ ($y=L$). GW  considered a bulk scalar in the above background with bulk action~\cite{Goldberger:1999uk}
\bea
S_B=\frac{1}{2}\int d^4x dy \sqrt{G}(G^{MN}\partial_M \phi\partial_N\phi -m^2\phi^2),
\eea 
and brane localized actions
\bea
S_{UV}&=& -\frac{1}{2}\int d^4x \sqrt{-g_{uv}}\bl (\phi^2-u^2)^2,\label{GW_S_UV}\\
S_{IR}&=& -\frac{1}{2}\int d^4x \sqrt{-g_{ir}}\lambda (\phi^2-v^2)^2,
\eea
where $g_{\mu\nu}^{uv}$ and $g_{\mu\nu}^{ir}$ are the restrictions of $G_{\mu\nu}$ to $y=0$ and $y=L$ respectively. Note that the boundary actions do not possess any odd terms in the field $\phi$ and consequently the entire action is invariant under a $Z_2$ symmetry $\phi\rightarrow-\phi$. 

We shall consider the case where the bulk and IR brane actions retain their form $S_B$ and $S_{IR}$ but modify the UV brane potential to\footnote{Note that the minimum of this UV potential is nonzero and will therefore contribute to the UV brane tension. In RS models the UV brane tension must be related to the bulk cosmological constant (CC) to ensure a vanishing 4D CC. In the presence of the UV action (\ref{example_uv_action}) one should shift the usual RS brane tension $V_{uv}\rightarrow V_{uv}-\Delta V_{uv}$ to cancel out this additional boundary tension and retain the standard RS solution. This constitutes a modified version of the usual tuning of the 4D CC present in RS models. It will be understood in the present work that this shift of the UV brane tension has been undertaken whenever the UV potential has a non-vanishing minimum so the usual RS background solution holds.}
\bea
S_{UV}&=& -\frac{1}{2}\int d^4x \sqrt{-g_{uv}}\bl (\phi^2+u^2)^2,\label{example_uv_action}
\eea
with $u^2>0$. Variation of $S_B$ produces the bulk equation of motion for $\phi$ as
\bea
\partial_N\left[\sqrt{G} G^{MN}\partial_M\phi\right] +\sqrt{G}m^2\phi=0,
\eea
which for $\partial_\mu \phi=0$ has the solution
\bea
\phi(y)=Ae^{\bp ky}+Be^{\bm ky}\label{gw_bulk_sol+}\;, 
\eea
where
\bea
\beta_{\pm}&=&2\pm\sqrt{4+\frac{m^2}{k^2}}\equiv 2\pm\nu.
\eea
The bulk solution must satisfy the following boundary conditions (BCs)
\bea 
& &\partial_y \phi(0) -2\bl (\phi^2(0)+u^2)\phi(0) =0,\\
& &\partial_y \phi(L) +2\lambda (\phi^2(L)-v^2)\phi(L) =0,
\eea
which, in the limit of large $\lambda,\bl$, are
\bea
\phi(0)&=&\delta\phi(0),\\
\phi(L)&=&v+\delta\phi(L),
\eea
where\footnote{We generically refer to quartic parameters by $\lambda$ so that $\mathcal{O}(\lambda^{-1})$ also stands for terms of $\mathcal{O}(\bl^{-1})$. This applies throughout the paper.} $\delta\phi=\mathcal{O}(\lambda^{-1})$. 

Momentarily ignoring the $\delta\phi$ corrections the leading order IR boundary contributions to the potential for the length of the extra dimension vanish and $V(L)$ can be written as
\bea
V(L)=V_B(L)+V_{UV},
\eea
where $V_{UV}= \bl u^4/2+\mathcal{O}(\lambda^{-1})$ and the bulk piece is
\begin{eqnarray}
V_B(L)=\frac{k}{2}\left[(\nu+2)A^2(e^{2\nu kL}-1)+(\nu-2)B^2 (1-e^{-2\nu kL})\right].
\eea
Enforcing the BCs on the bulk solution gives
\bea
A=-B\simeq ve^{-\bp kL},
\eea
and for $m^2/k^2<1$ we may follow~\cite{Goldberger:1999uk} and write $\nu= 2+\epsilon$ where $\epsilon \simeq m^2/4k^2$, giving
\bea
V(L)&=&2kA^2(e^{2\nu kL}-1) +\mathcal{O}(\epsilon),\\
&\simeq& 2kv^2e^{-4kL} + \ .\ .\ . \ .
\eea 
where the dots denote the constant piece and terms of $\mathcal{O}(\lambda^{-1})$ and $\mathcal{O}(\epsilon)$. This is the leading order potential $L$. We would like to also determine the $\mathcal{O}(\lambda^{-1})$ corrections to the potential. The corrections to the BCs are found to be
\bea
\delta\phi(0)&=&\frac{k}{2\bl u^2}(A\bp+B\bm)\simeq \frac{2k }{\bl u^2} e^{-(4+\epsilon) kL}v,\\
\delta\phi(L)&=&-\frac{k}{4\lambda v^2}(A\bp e^{\bp kL}+B\bm e^{\bm kL})\simeq -\frac{k}{\lambda v^2}v,
\eea
which give the following corrections to $A,B$:
\bea
\delta A&=& \frac{(\delta\phi(L)-\delta\phi(0)e^{\beta_{-} kL})}{e^{\bp kL}-e^{\bm kL}}\nonumber\\
&\simeq& -e^{-\bp kL} v\frac{k}{\lambda v^2},\\
\delta B&=& \frac{(\delta\phi(0)e^{\beta_{+} kL}-\delta\phi(L))}{e^{\bp kL}-e^{\bm kL}}\nonumber\\
&\simeq& e^{-\bp kL} v\left[\frac{k}{\lambda v^2}+\frac{2k}{\bl u^2}\right].  
\eea
This produces a correction to the bulk potential for $L$:
\bea
\delta V_B(L)&\simeq& -4kv^2e^{-4kL}\left[\frac{k}{\lambda v^2}\right]
+ \ .\ .\ .\ ,
\eea
and results in a nonzero contribution to $V(L)$ from the boundary potentials:
\bea
\delta V_{IR}(L)&\simeq&2kv^2e^{-4kL}\left[\frac{k}{\lambda v^2}\right] + \ .\ .\ .\ ,\\ 
 V_{UV}(L)&\simeq&\frac{\bl u^4}{2}+4kv^2e^{-4kL}\left[\frac{k}{\bl u^2}e^{-(4+2\epsilon)kL}\right] +\ .\ .\ .\ .
\eea
Putting all these results together the complete potential for $L$ through $\mathcal{O}(\lambda^{-1})$ is given by
\bea
V(L)&=&V_B+\delta V_B+\delta V_{IR}+ V_{UV}\nonumber\\
&\simeq&2ke^{-4kL}v^2 \left[1-\frac{k}{\lambda v^2}\right] + \frac{\bl
  u^4}{2}+\ .\ .\ .\ ,\label{VL_runaway_1throat}
\eea
the minimum of which corresponds to $L\rightarrow\infty$. Thus we learn that if the GW scalar has a potential on the UV brane whose minimum corresponds to a vanishing brane vacuum expectation value (VEV),
 the potential for the length of the extra dimension does not stabilize $L$ at a finite value. Instead the radius runs away and the extra dimension is not compactified. This feature will occur in some of the multithroat scenarios below.

If one instead uses the UV action (\ref{GW_S_UV}) as in~\cite{Goldberger:1999uk} the calculation carries through in the same fashion, however instead of (\ref{VL_runaway_1throat}) one obtains
\bea
V(L)&\simeq&2ke^{-4kL}(v-ue^{-\epsilon kL})^2 \left[1-\frac{k}{\lambda v^2}\right] + \ .\ .\ .\ .
\eea
This potential differs from (\ref{VL_runaway_1throat}) in an important way since the minimum is now at
\begin{eqnarray}
L=\frac{1}{\epsilon k}\ln\left(\frac{u}{v}\right),
\eea
and the extra dimension is stabilized at a finite value. This is the
GW result. Notice that values of $Lk\sim\mathcal{O}(10)$ are easily
obtained, an important feature that is necessary for the IR scale to
be naturally much less than the Planck scale, $M_{IR}\sim e^{-
  kL}M_{Pl}\ll M_{Pl}$, as required to solve the hierarchy
problem~\cite{Randall:1999ee}. 

For future reference we note that the observed failure of the GW
scalar to stabilize the extra dimension for a UV potential minimized
by $\phi(0)=0$ holds more generally for non-zero $\phi(0)=u$ if
$u<v$. If the energy density of the scalar profile is dominated in the
IR a runaway solution is prefered as increasing the size of the extra
dimension redshifts this energy. The stable solution therefore
requires $u/v>1$.

Following~\cite{Goldberger:1999uk} we have neglected the backreaction
of the GW scalar on the metric in the above analysis. This is
acceptable provided $u^2/M^3_*\ll1$ and $v^2/M_*^3\ll1$, where the 5D
gravity scale $M_*$ satisfies $M_{Pl}^2\simeq M_*^3/k$. The
backreaction has been considered for a modified version of the GW
analysis in~\cite{DeWolfe:1999cp,Csaki:2000zn} and more recently a
partial inclusion of the backreaction in the GW scenario has
appeared~\cite{Konstandin:2010cd}. Related analysis can also be found
in~\cite{Lesgourgues:2003mi}. We shall not consider the backreaction
in the generalizations that follow and one should keep in mind that
appropriate (and obvious) generalizations of the conditions
$u^2/M^3_*,v^2/M_*^3\ll1$ must also hold. Let us also note that a
dicussion on the 4D interpretation of the GW mechanism
based on AdS/CFT is given in~\cite{ArkaniHamed:2000ds} and
generalizations of the GW mehcanism for soft-wall models have appeared~\cite{Cabrer:2009we}.
%%%%%%%%%%%%%%%%%%%%%%%%%%%%%%%%%%%%%%%%%%%%%%%%%%%
\section{GW Mechanism for Two Independent Throats\label{sec:2ind_throats}} 
In this section we shall generalize the GW mechanism to models with
two independent warped throats glued together at a common UV brane. We 
label the two throats as $i=1,2,$ with the metric in the $i$-th throat defined by
\bea
ds^2_i=e^{-2ky_i}\eta_{\mu\nu}dx^\mu dx^\nu -dy_i^2\equiv G^i_{MN}dx^M_i dx^N_i,
\eea
where  $x^\mu_i\equiv x^\mu$ are the 4D coordinates and the warped
extra dimensions are labeled by $x_i^5\equiv y_i\in[0,L_i]$. The IR
branes are located at $y_i=L_i$ and the common UV brane sits at
$y_i=0$, $\forall i$. The gravitational sources necessary to realize
this background have been discussed in~\cite{Cacciapaglia:2006tg} and
we refer the reader there for details. In this section and throughout
we consider IR-UV-IR type constructs. The kinetic term for the
massless radion in such setups has the correct sign so the stability
issues associated with the UV-IR-UV setups are not present. Also note that we consider a common bulk cosmological constant for the two throats so that the curvature $k$ is the same in both throats.

We consider  a GW scalar $\phi_i$ in each throat with the bulk action in the $i$-th throat given by
\bea
S_B^i=\frac{1}{2}\int d^4x dy_i \sqrt{G^i}(G_i^{MN}\partial_M \phi_i\partial_N\phi_i -m_i^2\phi_i^2).
\eea 
The IR brane localized actions are
\bea
S^i_{IR}&=& -\frac{1}{2}\int d^4x \sqrt{-g^i_{ir}}\lambda_i (\phi_i^2-v_i^2)^2,
\eea
and the action on the common UV brane is
\bea
S_{UV}&=& -\frac{1}{2}\int d^4x \sqrt{-g_{uv}}\left\{\sum_i\bl_i(\phi_i^2-u_i^2)^2+\kappa \phi_1^2\phi_2^2\right\},
\eea
where $g_{\mu\nu}^{uv}$ and $(g_{\mu\nu}^{ir})^i$ are the restrictions of $G^i_{\mu\nu}$ to $y_i=0$ and $y_i=L_i$ respectively. The total scalar action is thus
\bea
S=\sum_i (S_B^i +S_{IR}^i)+S_{UV},
\eea
which does not possess any odd terms in the fields $\phi_i$ and is therefore invariant under two independent $Z_2$ symmetries:
\bea
Z_2^{(i)}:\ \phi_i\rightarrow-\phi_i.
\eea
These symmetries generalize the discrete symmetry of the GW mechanism and serve primarily to simplify the calculation.

Variation of $S_B^i$ produces the bulk equation of motion for $\phi_i$ in the $i$-th throat:
\bea
\partial_N\left[\sqrt{G^i} G_i^{MN}\partial_M\phi_i\right] +\sqrt{G^i}m^2_i\phi_i=0,
\eea
which for $\partial_\mu \phi_i=0$ has the solution
\bea
\phi_i(y_i)=A_ie^{\bp^i ky_i}+B_ie^{\bm^i ky_i},\label{gw_bulk_sol_i}
\eea
where
\bea
\beta_{\pm}^i&=&2\pm\sqrt{4+\frac{m_i^2}{k^2}}\equiv 2\pm\nu_i=2\pm (2+\epsilon_i),
\eea
and $\epsilon_i\simeq m_i^2/4k^2$ for $m_i<k$. The demand that the variation of the action vanish on the boundaries leads to the BCs:
\bea 
& &\partial_y \phi_i(0) -2\bl_i (\phi_i^2(0)-u_i^2)\phi_i(0)-\kappa\phi_j^2(0)\phi_i(0) =0,\quad i\ne j,\\
& &\partial_y \phi_i(L_i) +2\lambda_i (\phi_i^2(L_i)-v_i^2)\phi_i(L_i) =0.
\eea

The potential for the throat lengths $L_i$ is a sum of contributions from both the bulks and the branes and may be written as
\bea
V(L_1,L_2)=\sum_i \left[ V_B^i(L_i) +V_{IR}^i(L_i)\right]+ V_{UV}(L_1,L_2).
\eea
Inserting the solution (\ref{gw_bulk_sol_i}) into the $i$-th bulk action and integrating over the extra dimension determines the bulk contribution to the potential,
\begin{eqnarray}
V^i_B(L_i)=\frac{k}{2}\left[(\nu_i+2)A_i^2(e^{2\nu_i kL_i}-1)+(\nu_i-2)B_i^2 (1-e^{-2\nu_i kL_i})\right].
\eea
As in the GW case we consider large $\lambda,\bl$, so the leading order BCs are $\phi_i(0)=u_i$ and $\phi_i(L_i)=v_i$, giving
\bea
A_i& \simeq& e^{-\bp^i kL_i}v_i -u_i e^{-2\nu_i kL_i},\\
B_i&\simeq& u_i(1+e^{-2\nu_i kL_i})-v_i e^{-\bp^i kL_i},
\eea
so that
\bea
V_B^i(L_i)&=&2kA_i^2(e^{2\nu_i kL_i}-1) +\mathcal{O}(\epsilon_i),\\
&\simeq& 2k e^{-4 kL_i}(v_i-u_ie^{-\epsilon_i kL_i})^2  + \ .\ .\ . \ .
\eea 
To leading order the IR brane contributions vanish and the UV brane contribution is a constant so that $V(L_1,L_2)$ is minimized at
\begin{eqnarray}
L_i=\frac{1}{\epsilon_i k}\ln\left(\frac{u_i}{v_i}\right).
\eea
As one would expect this matches the GW result as we have effectively neglected the coupling term $\kappa\phi_1^2\phi_2^2$ on the UV to leading order. To find the corrections induced by $\kappa$ we write
\bea
\phi_i(0)&=&u_i+\delta\phi_i(0),\\
\phi_i(L_i)&=&v_i+\delta\phi_i(L_i),
\eea
and find that
\bea
\delta\phi_i(L_i)&\simeq& -\frac{k}{\lambda_i v_i^2}(v_i-u_ie^{-\epsilon_i kL_i}),
\eea
and
\bea
\delta\phi_i(0)&\simeq&\frac{k}{\bl_i u_i^2}\left[(v_i-u_ie^{-\epsilon_i kL_i})e^{-\bp^i kL_i}- \frac{\kappa}{4k} 
u_j^2u_i\right],\quad i\ne j. 
\eea

Enforcing the corrected BCs on the bulk solution results in the following corrections to $A,B$:
\bea
\delta A_i&\simeq& -e^{-\bp^i kL_i} \left\{(v_i-u_ie^{-\epsilon_i kL_i})\frac{k}{\lambda_i v_i^2}-\frac{\kappa}{4}\frac{u_j^2}{\bl_i u_i}e^{-\epsilon_i k L_i}\right\},\quad i\ne j,
\\
\delta B_i&\simeq& e^{-\bp^i kL_i}(v_i-u_ie^{-\epsilon_i kL_i})\left[\frac{k}{\lambda_i v_i^2}+\frac{k}{\bl_i u_i^2}\right]-\frac{\kappa}{4}\frac{u_j^2}{\bl_i u_i}
,\quad i\ne j.
\eea
With these results we can calculate the $\mathcal{O}(\lambda^{-1})$ corrections to $V(L_1,L_2)$. These include corrections to the bulk potentials ($\delta V^i_B$), the IR boundary potentials ($\delta V^i_{IR}$) and the UV potential, giving
\bea
V(L_1,L_2)&=&\sum_i \left[V^i_B+ \delta V^i_B+\delta V^i_{IR}\right]+V_{UV}\nonumber\\
&\simeq& \sum _i 2 k e^{-4kL_i}(v_i-u_i e^{-\epsilon_i kL_i})\times\nonumber\\
& &\quad \quad \left\{(v_i-u_i e^{-\epsilon_i kL_i})\left(1-\frac{k}{\lambda_i v_i^2}\right) +\frac{\kappa u_j^2}{2\bl_i u_i}e^{-\epsilon_ikL_i}\right\}+ \ .\ .\ . \ , \quad j\ne i, 
\eea
where the dots denote subdominant terms and a constant piece. To leading order in $\lambda^{-1}$ the minimum of $V(L_1,L_2)$ is given by
\begin{eqnarray}
L_i=\frac{1}{\epsilon_i k}\left\{\ln\left(\frac{u_i}{v_i}\right)+
\ln\left(1-\frac{\kappa u_j^2 v_i^2 \lambda_i}{2\bl_i u_i^2 (v_i^2 \lambda_i - k)}\right)\right\}
, \quad \quad j\ne i.\label{2_throat_ind_length}
\eea
Thus the presence of independent GW scalars in each throat, with a common UV brane coupling, generates a potential for the throat lengths whose minimum is set by finite values of $L_{1,2}$. As in GW one may obtain $L_i k \simeq \mathcal{O}(10)$ without fine tuning. The minimizing throat lengths return to the usual GW result in each throat in the limit $\kappa\rightarrow 0$, which is to be expected as the stabilization dynamics of the two throats decouple in this limit. For finite $\kappa$ there is a correction to the GW result.

 In general the lengths (\ref{2_throat_ind_length}) are expected to differ in each throat. As such this generalization of the GW method is useful for models with two independent throats with distinct IR scales, as in~\cite{Gripaios:2006dc} and~\cite{McDonald:2010iq} in which the SM resides in one throat with the usual order TeV IR scale and a hidden sector resides in a second throat with an independent IR scale. Ref.~\cite{McDonald:2010iq} employed a hidden IR scale of order a GeV, which corresponds $L_1/L_2\sim0.8$, and such a difference is easily obtained with Eq. (\ref{2_throat_ind_length}). Our results could also be employed in the scenario of~\cite{Cacciapaglia:2005pa} in which the third generation is sequestered in a separate throat with an independent IR scale relative to that in which the lighter generations reside.
%%%%%%%%%%%%%%%%%%%%%%%%%%%%%%%%%%%%%%%%%%%%%%%%%%%
\section{GW Mechanism for Two Throats with a $\mathbf{Z_2}$\label{sec:GW_z2}}
We would like to consider the interesting case where the two throats are related by a $Z_2$ interchange symmetry. Such a gravitational background has already been employed in the literature~\cite{Agashe:2007jb} and is of interest because the interchange symmetry can affect dynamics and motivate phenomenological applications. The action of the interchange symmetry is:
\bea
Z_2:\quad y_1\leftrightarrow y_2,
\eea
with a corresponding action on the field content of each throat:
\bea
 \mathcal{F}_1(x^\mu, y_1)&\leftrightarrow& \mathcal{F}_2(x^\mu,y_2).
\eea
The fields $\mathcal{F}_i$ could denote the wavefunction in the $i$-th throat of a field that propagates in both throats; an example being a SM gauge boson propagating in two throats that are subject to a ``UED-like''~\cite{Appelquist:2000nn} reflection parity. This scenario was considered in~\cite{Agashe:2007jb} where it was shown that the resulting KK-parity ensures stability of the lightest odd KK mode, thereby motivating a dark matter candidate for RS models. Alternatively the SM may be confined to one throat with a hidden sector residing in the other, as would occur in a multithroat version of the mirror matter models~\cite{Foot:1991bp,Berezhiani:1995yi,Okun:2006eb}. Our interests here are primarily in the generalization of the GW mechanism to such a symmetric throat arrangement, regardless of the specific application.

We label the throats as $i=1,2,$ with the metric in the $i$-th throat is again defined by 
\bea
ds^2_i=e^{-2ky_i}\eta_{\mu\nu}dx^\mu dx^\nu -dy_i^2\equiv
G^i_{MN}dx^M_i dx^N_i,  
\label{mulithroat_metric}
\eea
and we consider  a GW scalar $\phi_i$ in each throat with the action of the interchange symmetry being
\bea
Z_2:\quad \phi_1\leftrightarrow \phi_2.
\eea
The bulk action in the $i$-th throat is
\bea
S_B^i=\frac{1}{2}\int d^4x dy_i \sqrt{G^i}(G_i^{MN}\partial_M \phi_i\partial_N\phi_i -m^2\phi_i^2),\label{bulk_action_z2}
\eea 
and the IR brane localized actions are 
\bea
S^i_{IR}&=& -\frac{1}{2}\int d^4x \sqrt{-g^i_{ir}}\lambda (\phi_i^2-v^2)^2,\label{ir_action_z2}
\eea
where $(g_{\mu\nu}^{ir})^i$ is the restriction of $G^i_{\mu\nu}$ to $y_i=L_i$. Note that the interchange symmetry requires equality of the bulk masses $m$ and the IR brane parameters $\lambda$ and $v$ for scalars in  distinct throats. For $\partial_\mu \phi_i=0$ the solution to the bulk equations of motion are
\bea
\phi_i(y_i)=A_ie^{\bp ky_i}+B_ie^{\bm ky_i},\label{gw_bulk_sol_i_z2}
\eea
where $\beta_{\pm}$ is defined as
\bea
\beta_{\pm}&=&2\pm\sqrt{4+\frac{m^2}{k^2}}\equiv 2\pm\nu=2\pm (2+\epsilon),
\eea
with $\epsilon\simeq m^2/4k^2$. The IR BC is
\bea 
& &\partial_y \phi_i(L_i) +2\lambda (\phi_i(L_i)^2-v^2)\phi_i(L_i) =0,
\eea
and we consider large $\lambda$ so that
\bea
\phi_i(L_i)=v,\quad\quad i=1,2,
\eea 
to leading order. We consider three distinct cases for the UV brane potential in what follows. Each case has different consequences for the structure of the resulting gravitational background.
%%%%%%%%%%%%%%%%%%%%%%%%%%%%%%%%%%%%%%%%%%%%%%%%%%%%%%%%%%%%
\subsection{Preservation of the $\mathbf{Z_2}$ Symmetry\label{preserving_z2}}
First we shall present a generalized GW mechanism to stabilize the two throats while preserving the $Z_2$ interchange symmetry. In this case we write the action on the common UV brane as
\bea
S_{UV}&=& -\frac{1}{2}\int d^4x \sqrt{-g_{uv}}\left\{\lp(\phi_1^2+\phi_2^2-2u^2)^2+\lm (\phi_1^2-\phi_2^2)^2\right\},\label{z2_pres_suv}
\eea
where $g_{\mu\nu}^{uv}$ is the restriction of $G^i_{\mu\nu}$ to $y_i=0$. As in the previous section the entire action is invariant under two independent $Z_2$ symmetries whose actions are defined by:
\bea
Z_2^{(i)}:\ \phi_i\rightarrow-\phi_i.
\eea
The corresponding UV BC is
\bea 
& &\left.[\partial_y \phi_i -2\lp (\phi_i^2+\phi_j^2-2u^2)
\phi_i-2\lm(\phi_i^2-\phi_j^2)\phi_i]\right|_{UV} =0,\quad i\ne j,
\eea
and for large $\lambda_{\pm}>0$ the leading order UV BCs are
\bea
\phi_i(0)=u,\quad\quad i=1,2.
\eea
Imposing the BCs on the bulk solutions and integrating out the extra dimension generates the following potential for $L_{1,2}$:
\bea
V(L_1,L_2)&\simeq&\sum_i 2k e^{-4 kL_i}(v-ue^{-\epsilon kL_i})^2  + \ .\ .\ . \ , 
\label{leading_z2_otential_pres}
\eea 
which is minimized at the usual GW value:
\begin{eqnarray}
L_1=L_2=\frac{1}{\epsilon k}\ln\left(\frac{u}{v}\right).\label{z2_pres_throat_l}
\eea
Equality of $L_{1}$ and $L_{2}$ implies preservation the $Z_2$ interchange symmetry in the presence of the stabilizing dynamics. One might wonder however if the $\mathcal{O}(\lambda^{-1})$ corrections modify this equality; inclusion of these corrections gives:
\bea
V(L_1,L_2)&=&\sum_i \left[V^i_B+ \delta V^i_B+\delta V^i_{IR}\right]+V_{UV}\nonumber\\
&\simeq& \sum _i2ke^{-4kL_i}(v-ue^{-\epsilon kL_i})^2
\left[1-\frac{k}{\lambda v^2}\right] + \ .\ .\ .\ .\ , \label{corrected_z2_otential_pres}
\eea
and the minimum remains at (\ref{z2_pres_throat_l}). 

This type of stabilization mechanism, in which an initially $Z_2$ symmetric throat configuration remains $Z_2$ symmetric as the GW scalars reach their minimum and fix the value of the throat lengths, can be employed in, e.g., the warped dark matter model of~\cite{Agashe:2007jb}. As the $Z_2$ symmetry is preserved any field properties dependent on this symmetry, like the stability of the dark matter candidate, remain in tact. 
%%%%%%%%%%%%%%%%%%%%%%%%%%%%%%%%%%%%%%%%%%%%%%%%%%%%%%%%%%%%
\subsection{Breaking the $\mathbf{Z_2}$ Symmetry\label{break_z2_inft}}
In this section we write the action on the common UV brane as\footnote{This is a simple rewriting of the potential used in Section~\ref{preserving_z2} with the parameters $\lambda_{a,b}$ and $u$ related to the parameters used in that section; see~\cite{Foot:2000tp} and~\cite{Barbieri:2005ri} for a discussion of a related potential.}
\bea
S_{UV}&=& -\frac{1}{2}\int d^4x \sqrt{-g^i_{uv}}\left\{\la(\phi_1^2+\phi_2^2-u^2)^2+\lb\phi_1^2\phi_2^2\right\},
\eea
which is also invariant under the two independent symmetries $\phi_i\rightarrow-\phi_i$. The UV BC is
\bea 
& &\left.[\partial_y \phi_i -2\la (\phi_i^2+\phi_j^2-u^2)
\phi_i-\lb\phi_j^2\phi_i]\right|_{UV} =0,\quad i\ne j.
\eea
For large $\lambda_{a,b}>0$ the leading order UV BCs correspond to just one scalar acquiring a boundary VEV,
\bea
\phi_1(0)=u,\quad\phi_2(0)=0,
\eea
where we label the field with a non-zero boundary VEV as $i=1$. Imposing the BCs gives
\bea
A_1& \simeq& ve^{-\bp kL_1} -u e^{-2\nu kL_1},\nonumber\\
B_1&\simeq& u(1+e^{-2\nu kL_1})-v e^{-\bp kL_1},\nonumber\\
A_2&=&-B_2 \simeq e^{-\bp kL_2} v
\eea
and to leading order the potential for $L_{i}$ is 
\bea
V_B^i&\simeq& 2k\times \left\{\begin{array}{ll} e^{-4
      kL_1}(v-ue^{-\epsilon kL_1})^2 +\ .\ .\ .\ ,& i=1\\ e^{-4 kL_2}v^2
    +\ .\ .\ .\ , & i=2\end{array}\right..
\eea 
Including the  $\mathcal{O}(\lambda^{-1})$ corrections the full potential is
\bea
V(L_1,L_2)&=&\sum_i \left[V^i_B+ \delta V^i_B+\delta V^i_{IR}\right]+ V_{UV}\nonumber\\
&\simeq& 2ke^{-4kL_1}(v-ue^{-\epsilon kL_1})^2 \left[1-\frac{k}{\lambda v^2}\right] \nonumber\\
& &\qquad\quad +\ 2k e^{-4kL_2}v^2 \left[1-\frac{k}{\lambda
    v^2}\right]+ \ .\ .\ .\  , 
\eea
yielding the following minimizing throat lengths:
\bea 
L_1=\frac{1}{\epsilon k}\ln\left(\frac{u}{v}\right) \quad\mathrm{and}\quad L_2\rightarrow \infty.
\eea
Observe that the $Z_2$ symmetry has been broken as $L_1\ne L_2$,
however only $L_1$ remains finite with $L_2$ running away to
infinity. This runaway is consistent with our example calculation for a single
throat in Sec.~\ref{sec:GW_example}.
%%%%%%%%%%%%%%%%%%%%%%%%%%%%%%%%%%%%%%%%%%%%%%%%%%%%%%%%%%%%
\subsection{Breaking the $\mathbf{Z_2}$ Symmetry with Finite Throat Lengths\label{sec:break_z2_finite}}
We would like to find a stabilization configuration that breaks the $Z_2$ interchange symmetry of the two throats and yet stabilizes both throats at finite lengths. In the preceding sections the scalar action possessed two additional discrete symmetries, $Z_2^{(i)}:\phi_i\rightarrow -\phi_i$, which generalize the discrete symmetry employed in~\cite{Goldberger:1999uk}. These symmetries prove too restrictive if one seeks to stabilize both throats at finite lengths as the asymmetric minimum of the resulting UV potential induces a non-zero boundary VEV for only one of the scalars. As discussed in Sec.~\ref{sec:GW_example} for the GW scalar, the UV VEV must dominate the IR VEV in order to stabilize the throat length. In this section we relax the symmetries $Z_2^{(i)}$ to permit an asymmetric minimum to the UV potential which permits both scalars to take non-zero VEVs. 

We employ the following UV action:
\bea
S_{UV}&=& -\frac{1}{2}\int d^4x \sqrt{-g_{uv}}\left\{\la(\phi_1^2+\phi_2^2-u^2)^2+\lb\phi_1^2\phi_2^2-\mu\phi_1\phi_2 -\kappa(\phi_1^3\phi_2+\phi_1 \phi_2^3)\right\}\label{asymmetric_z2_potential}
\eea
where the $\mu$ and $\kappa$ terms break the symmetries $Z_2^{(i)}$ but admit the following diagonal symmetry:
\bea
Z_2^{D}:\quad \phi_{1,2}\rightarrow -\phi_{1,2}.
\eea
Retaining (\ref{bulk_action_z2}) and (\ref{ir_action_z2}) for the bulk and IR actions respectively we see that $Z_2^D$ is a symmetry of the entire action. 

With the above UV action one obtains the  following  BCs:
\bea 
& &\left.[\partial_y \phi_i -2\la (\phi_i^2+\phi_j^2-u^2) 
\phi_i-\lb\phi_j^2\phi_i +\frac{\mu}{2}\phi_j +\frac{\kappa}{2}\phi_j(\phi_j^2+3\phi_i^2)]\right|_{UV}=0,\quad i\ne j.\nonumber
\eea
We again consider the limit where the coupling constants in the UV Lagrangian are large and the derivative piece is subdominant. However to simplify the calculation we consider the case where the  $\lambda_{a,b}$ terms also dominate the $\mu$ and $\kappa$ terms. This hierarchy of parameters\footnote{As $\lambda_{a,b}$ and $\kappa$ have different mass dimension to $\mu$ this statement must be made with reference to dimensionless quantities by employing appropriate powers of a fixed reference scale like the curvature.} is technically natural as the symmetry of the action is enhanced from $Z_2^D$ to $Z_2^{(1)}\times Z_2^{(2)}$ in the limit $\mu,\kappa\rightarrow0$. We arrive at the following leading order BCs
\bea
\phi_1(0)=u,\quad\quad \phi_2(0) = \frac{\mu +\kappa u^2}{2\lambda_b u^2}u,
\eea
where terms of order $\lambda^{-2}$ are neglected
\footnote{We have checked that this critical point of the UV potential is
indeed stable with $\lambda_{a,b} \gg \mu, \kappa$ and $\lambda_{a,b}
>0$.}. 

Based on what we have seen in the preceding sections we can immediately deduce that the leading order potential for the throat lengths is
\bea
V_B(L_1,L_2)&\simeq& 2k e^{-4 kL_1}(v-ue^{-\epsilon kL_1})^2 +2k e^{-4 kL_2}(v-ue^{-\epsilon kL_2}\times \frac{\mu +\kappa u^2}{2\lambda_b u^2})^2 +\ .\ .\ . \ ,\nonumber
\eea 
minimization of which  gives
\bea 
L_1&=&\frac{1}{\epsilon k}\ln\left(\frac{u}{v}\right),\\
L_2&=&\frac{1}{\epsilon k}\left\{\ln\left(\frac{u}{v}\right)+\ln\left(\frac{\mu+\kappa u^2}{2\lb u^2}\right)\right\}.
\eea
 Thus the throat interchange symmetry is broken by the stabilization dynamics and both throats acquire finite lengths. With $\lb,\mu,\kappa>0$ the second logarithm is negative and one has $L_2<L_1$. Including the $\mathcal{O}(\lambda^{-1})$ corrections from the derivative pieces in the BCs gives 
\bea
V(L_1,L_2)&=&\sum_i \left[V^i_B+ \delta V^i_B+\delta V^i_{IR}\right]+V_{UV}\nonumber\\
&\simeq& 2ke^{-4kL_1}(v-ue^{-\epsilon kL_1})^2 \left[1-\frac{k}{\lambda v^2}\right] \nonumber\\
& &\qquad\quad +\ 2k e^{-4kL_2}\left\{v-ue^{-\epsilon kL_2}\times \frac{\mu +\kappa u^2}{2\lambda_b u^2}\right\}^2\left[1-\frac{k}{\lambda v^2}\right]+ \ .\ .\ .\ ,
\eea
and the leading order expressions for $L_{1,2}$ hold. This shows that the generalized GW mechanism can successfully  break the interchange symmetry and fix the throat lengths at finite values. Such a mechanism would be of use in a flavor model based on a  broken $Z_2$ symmetry or in a warped realization of the broken mirror model discussed in~\cite{Berezhiani:1995yi}. 

Note that there is a sense in which symmetry breaking occurs both in the UV (different GW VEVs) and in the IR (different throat lengths) in the present example. Ultimately it is the UV behaviour of the GW scalars that triggers the symmetry breaking, which then manifests in the IR in the form of distinct IR scales. However it is interesting that the emergence of multiple energy scales in the IR encodes information about the UV.

It is also important to make some comments on the above solution. Stability of the solution requires the UV boundary VEV to be larger than the IR VEV and therefore the solution in the second throat necessitates
\bea
 \left(\frac{\mu +\kappa u^2}{2\lambda_b u^2}\right) >\frac{v}{u}.
\eea
Accordingly one cannot take $(\mu +\kappa u^2)/(2\lambda_b u^2)$ to be
``too small''  as the limit $\lb\rightarrow\infty$ corresponds to
$\phi_2(0)=0$ and results in the runaway solution
$L_2\rightarrow\infty$ found in the previous subsection. If one
demands that $u/v$ does not greatly exceed $10$ based on naturalness
arguments then there is only a small window of parameter space in
which our results hold and our approximations can be trusted. More
generally there exists parameter space such that the asymmetric minima of the UV boundary potential
in Eq. (\ref{asymmetric_z2_potential}) has both $\phi_1(0)\sim u$ and $\phi_2(0)\sim u$. These solutions require that the $Z_2^{(i)}$ symmetry breaking terms are not subdominant. To see this note that in the limit $\kappa=0$ the minima of the UV boundary potential must satisfy
\bea
\left.\phi_1\phi_2\right|_{UV}=\frac{\mu}{2\lb},
\eea
where no approximation has been made. Furthermore the parameters must
also satisfy $\mu/\lb u^2 <1$. Therefore the solutions with both $\phi_1(0)\sim u$ and $\phi_2(0)\sim u$ require $(\mu/\lb u^2)\sim 1$ and lie outside our approximation. There is no difficulty of principle in employing these solutions however the calculation becomes considerably more cumbersome. For this reason we have considered the above approximations which in any case demonstrate an ability to stabilize the throat lengths via $Z_2$ symmetry breaking dynamics.

Before concluding this section we note that one can break the $Z_2$ interchange symmetry without introducing $Z_2^{(i)}$ breaking terms at the expense of an extended field content. Introducing a UV localized scalar $\chi$ that is odd under $Z_2$ permits a UV term $\propto \chi (\phi_1^2-\phi_2^2)$. If $\chi$ develops a VEV the $Z_2$ symmetry is broken and $\phi_{1,2}$ acquire distinct UV boundary values resulting in distinct finite throat lengths.
%%%%%%%%%%%%%%%%%%%%%%%%%%%%%%%%%%%%%%%%%%%%%%%%%%%
\section{GW Mechanism for Three Throats with a $\mathbf{Z_3}$\label{sec:GW_z3}}
Next we  consider three throats related by a cyclic $Z_3$ interchange symmetry, the action of which is given by:
\bea
Z_3:\quad y_i\rightarrow y_{i+1},
\eea
where $i=1,2,3,$ labels the throats, $y_i$ labels the warped extra dimension in each throat and $i+1$ is defined mod $3$ so that $y_3\rightarrow y_1$. The metric in the $i$-th throat is defined as per Eq. (\ref{mulithroat_metric}) and the action of $Z_3$ on the field content of each throat is:
\bea
 \mathcal{F}_i(x^\mu, y_i)&\rightarrow& \mathcal{F}_{i+1}(x^\mu,y_{i+1}).
\eea
To generalize the approach of GW to the three throat background we consider a set of GW scalars $\phi_i$, one in each throat, that  transform under $Z_3$ as $\phi_i\rightarrow \phi_{i+1}$. The bulk action in the $i$-th throat is
\bea
S_B^i=\frac{1}{2}\int d^4x dy_i \sqrt{G^i}(G_i^{MN}\partial_M \phi_i\partial_N\phi_i -m^2\phi_i^2),\label{bulk_action_z3}
\eea 
where mass equality is enforced by the $Z_3$ symmetry. The IR brane localized actions take the standard form
\bea
S^i_{IR}&=& -\frac{1}{2}\int d^4x \sqrt{-g^i_{ir}}\lambda (\phi_i^2-v^2)^2,\label{ir_action_z3}
\eea
where $(g_{\mu\nu}^{ir})^i$ is the restriction of $G^i_{\mu\nu}$ to $y_i=L_i$ and equality of the constants $\lambda$ and $v$ on the different branes is again dictated by symmetry. With these actions the IR BC in the $i$-th throat is
\bea 
& &\left.\left[\partial_y \phi_i +2\lambda (\phi_i^2-v^2)\phi_i\right]\right|_{IR} =0.
\eea
For large $\lambda$ one has $\phi_i(L_i)=v$ to leading order. A determination of the potential for $L_i$ requires specification of the UV action. In what follows we first consider a UV action that preserves the $Z_3$ symmetry and then present others that break it.
%%%%%%%%%%%%%%%%%%%%%%%%%%%%%%%%%%%%%%%%%%%%%%%%%%%%%%%%%%%%
\subsection{Preservation of the $\mathbf{Z_3}$ Symmetry\label{preserviing_z3}}
In this section we present a generalized GW mechanism for three throats subject to a $Z_3$ interchange symmetry that stabilizes the throats with identical lengths and therefore preserves the $Z_3$ symmetry. Applications for a symmetric three throat configuration have been discussed in~\cite{Cacciapaglia:2006tg}. These include a discrete family symmetry based on the interchange of three identical throats and a geometric realization of the trinification model~\cite{Carone:2004rp}. The $Z_3$ preserving system presented here provides an appropriate gravitational background for the examples discussed in~\cite{Cacciapaglia:2006tg}.
 
To obtain a $Z_3$ preserving configuration we employ the following UV action:
\bea
S_{UV}&=& -\frac{1}{2}\int d^4x \sqrt{-g_{uv}}\left\{-\mu^2 \sum_i\phi_i^2+\bl\left[\sum_i\phi_i^2\right]^2+\tl \sum_i\phi_i^4\right\},
\eea
where $g_{\mu\nu}^{uv}$ is the restriction of $G_{\mu\nu}$ to $y_i=0$. As this  action does not contain any odd terms in the fields $\phi_i$ the entire action is invariant under three independent $Z_2$ symmetries whose actions are defined by
\bea
Z_2^{(i)}:\ \phi_i\rightarrow-\phi_i.
\eea
These symmetries generalize the discrete $\phi\rightarrow -\phi$ symmetry of~\cite{Goldberger:1999uk}.

 Demanding that the variation of the action  vanishes on the UV brane gives the UV BC:
\bea 
& &\left[\partial_y \phi_i +\mu^2\phi_i -2\bl \left(\sum_j \phi_j^2\right)\phi_i -2\tl\phi_i^3\right]_{UV}=0.
\eea
For large $\bl,\ \tl,\ \mu>0$ the leading order UV BCs are the same for each scalar:
\bea
\phi_i^2(0)=\frac{\mu^2}{2(3\bl +\tl)}\equiv u^2,\quad\quad i=1,2,3.
\eea
Imposing the BCs and calculating the potential gives:
\bea
V(\{L_i\})&=&\sum_i 2kA_i^2(e^{2\nu kL_i}-1) +\mathcal{O}(\epsilon),\\
&\simeq&\sum_i 2k e^{-4 kL_i}(v-ue^{-\epsilon kL_i})^2  + \ .\ .\ .\ ,
\eea 
which is minimized at the usual GW value:
\begin{eqnarray}
L_i=\frac{1}{\epsilon k}\ln\left(\frac{u}{v}\right)\quad,\quad i=1,2,3.\label{throat_length_equal_z3}
\eea
Equality of $L_{1}$, $L_{2}$ and $L_3$ preserves the $Z_3$ interchange symmetry to leading order. Including the $\mathcal{O}(\lambda^{-1})$ corrections modifies the potential to:
\bea
V(\{L_i\})&=&\sum_i \left[V^i_B+ \delta V^i_B+\delta V^i_{IR}\right]+V_{UV}\nonumber\\
&\simeq& \sum _i2ke^{-4kL_i}(v-ue^{-\epsilon kL_i})^2 \left[1-\frac{k}{\lambda v^2}\right] + \ .\ .\ .\ , 
\eea
but the minimum remains at (\ref{throat_length_equal_z3}) and the $Z_3$ symmetry preserved.
%%%%%%%%%%%%%%%%%%%%%%%%%%%%%%%%%%%%%%%%%%%%%%%%%%%
\subsection{Breaking the $\mathbf{Z_3}$ Symmetry }
For alternative applications of a three throat background one may prefer a $Z_3$ symmetric action to produce three throats with different lengths as a result of stabilization; see~\cite{Abel:2010kw} for example. We can achieve this by rewriting the UV action as
\bea
S_{UV}&=& -\frac{1}{2}\int d^4x \sqrt{-g_{uv}}\left\{\la(\sum_i \phi_i^2-u^2)^2+\lb(\phi_1^2\phi_2^2+\phi_2^2\phi_3^2+\phi_3^2\phi_1^2)\right\},\nonumber\label{uv_action_z3_to_z2}
\eea
 and taking $\lambda_{a,b}>0$. The absence of odd terms in the fields $\phi_i$ ensures that the symmetries $Z_2^{(i)}$ still hold. For large $\lambda_{a,b}>0$ the leading order UV  BCs that minimize the UV potential are
\bea
\phi_1(0)=u\quad,\quad \phi_2(0)=\phi_3(0)=0,
\eea
which break the $Z_3$ symmetry and stabilize one throat (which we label as $i=1$) at 
\begin{eqnarray}
L_1=\frac{1}{\epsilon k}\ln\left(\frac{u}{v}\right).
\eea
However the vanishing UV boundary values for $\phi_{2,3}$ to leading order result in the $i=2$ and $i=3$ throat lengths running away, $L_{2,3}\rightarrow \infty$. This is similar to the broken $Z_2$ symmetry case of Section~\ref{break_z2_inft}. Although this successfully breaks the $Z_3$ interchange symmetry of the throats the runaway nature of the $i=2,3$ throats means such a solution is likely of limited phenomenological utility. 

In order to stabilize $L_{2,3}$ at finite values one should introduce $Z^{(i)}_2$ breaking terms in the UV action. A suitable action is:
\bea
S_{UV}&=& -\frac{1}{2}\int d^4x \sqrt{-g^i_{uv}}\left\{\la(\sum_i \phi_i^2-u^2)^2+\lb(\phi_1^2\phi_2^2+\phi_2^2\phi_3^2+\phi_3^2\phi_1^2)\right.\nonumber\\
& &-\mu (\phi_1\phi_2+\phi_2\phi_3+\phi_3\phi_1) -\kappa (\phi_1\phi_2\phi_3^2+\phi_2\phi_3\phi_1^2+\phi_3\phi_1\phi_2^3),\nonumber\\
& &\left.-\alpha (\phi_1\phi_3^3+\phi_2\phi_1^3+\phi_3\phi_2^3)-\beta(\phi_1\phi_2^3+\phi_2\phi_3^3+\phi_3\phi_1^3) \frac{}{}\right\}.
\eea
where the terms with coefficients $\mu,\ \kappa,\ \alpha$ and $\beta$ break the symmetries $Z_2^{(i)}$ but do not break the throat interchange symmetry $Z_3$. The UV action remains invariant under the diagonal discrete symmetry,
\bea
Z_2^D:\quad \phi_{1,2,3}\rightarrow-\phi_{1,2,3} \;,
\eea
which is also preserved by the bulk and IR actions. With this action the UV BCs are 
\bea 
& &\left[\partial_y \phi_i -\left\{2\la(\sum_\ell \phi_\ell^2-u^2)+\lb(\phi_j^2+\phi_k^2)\right\}\phi_i+\frac{\mu}{2}(\phi_j+\phi_k)+\right.\nonumber\\
& &\left.\qquad  \frac{\kappa}{2}(\phi_j\phi_k^2+\phi_k\phi_j^2+2\phi_i\phi_j\phi_k)+\frac{\alpha}{2}(\phi_k^3+3\phi_j\phi_i^2)+\frac{\beta}{2}(\phi_j^3+3\phi_k\phi_i^2)\right]_{UV}=0,
\eea
where $ i\ne j\ne k\ne i$. For computational simplicity we consider the limit where the boundary potential terms dominate the derivative piece in this BC. We further take the $\lambda_{a,b}$ terms to be larger than the $Z_2^{(i)}$ symmetry breaking terms, a technically natural limit. In this case the leading order BCs are\footnote{Like in the $Z_2$ case, the Hessian analysis indicates that this set
of solutions corresponds to a stable critical point as long as
$\lambda_{a,b} >0$ is dominant.} 
\bea
\phi_1(0)&=&u,\nonumber\\
\phi_2(0)&=&\frac{\mu+\alpha u^2}{2\lb u^2}u\equiv u_2,\nonumber\\
\phi_3(0)&=&\frac{\mu+\beta u^2}{2\lb u^2}u\equiv u_3,
\eea
and to leading order the potential for $L_{1,2,3}$ is 
\bea
V(\{L_i\})&=&\sum_i 2kA_i^2(e^{2\nu kL_i}-1) +\mathcal{O}(\epsilon),\\
&\simeq& 2k e^{-4 kL_1}(v-ue^{-\epsilon kL_1})^2+\sum_{i=2,3} 2k e^{-4 kL_i}(v-u_ie^{-\epsilon kL_i})^2  + \ .\ .\ . \ \label{break_z3_finite_zeroth}.
\eea 
The minimum of this potential occurs at
\begin{eqnarray}
L_1&=&\frac{1}{\epsilon k}\ln\left(\frac{u}{v}\right),\nonumber\\
L_2&=&\frac{1}{\epsilon k}\left\{\ln\left(\frac{u}{v}\right)+\ln\left(\frac{\mu+\alpha u^2}{2\lb u^2}\right)\right\},\nonumber\\
L_2&=&\frac{1}{\epsilon k}\left\{\ln\left(\frac{u}{v}\right)+\ln\left(\frac{\mu+\beta u^2}{2\lb u^2}\right)\right\}.\label{throat_length_z3_broke}
\eea
Note that all three throat lengths are finite and in general $L_i\ne L_j$ for all $ i\ne j$. Including the $\mathcal{O}(\lambda^{-1})$ corrections from the derivative pieces in the UV BCs the leading terms of Eq. (\ref{break_z3_finite_zeroth}) pick up a factor of $(1- k/\lambda v^2)$ %[SD]%
and the minimum agrees with (\ref{throat_length_z3_broke}) to $\mathcal{O}(\lambda^{-1})$.

The fact that all three throat lengths  are different reflects the
complete breaking of the $Z_3$ interchange symmetry. Such a GW
scenario is relevant for, e.g., the work of~\cite{Abel:2010kw} in
which a three throat configuration with a $Z_3$ interchange symmetry
was considered. In that work each generation of SM fermions was
confined to a different throat so the interchange symmetry of the
throats also served as a flavor symmetry. The $Z_3$ flavor symmetry
was posited to be broken by unspecified dynamics that result in
different throat lengths. The calculations of this section provide a
concrete realization of the gravitational background employed in that
work. We note that the $Z_3$ symmetry breaking structure obtained here
is not precisely
that envisioned in~\cite{Abel:2010kw}. While they sought to have the
$Z_3$ symmetry broken by different throat lengths they also sought to
have the UV brane remain $Z_3$ symmetric. Though our approach differs, the UV symmetry
breaking inherent in our methodology may not significantly alter
their conclusions. As the GW scalars are odd under $Z_2^D$ the coupling of SM
fields to the $Z_3$ breaking parameters may be sufficiently
sequestered to retain the $Z_3$ flavor symmetry to good approximation
in the UV, provided
the SM fermions are not charged under this symmetry. It would be interesting to consider alternative approaches
to determine if a purely symmetric UV sector can be found in
conjunction with $Z_3$ breaking in the IR. We also note that ideas
similar to those discussed here may be relevant to the three throat configuration of~\cite{Bechinger:2009qk}.

The comments made at the end of Sec.~\ref{sec:break_z2_finite} for the broken $Z_2$ case also apply here. It is important that one does not take the arguments of the second logarithms in the expressions for $L_{2,3}$ (\ref{throat_length_z3_broke}) to be ``too small'' compared to $v/u$ as one returns to the runaway solutions $L_{2,3}\rightarrow\infty$ in the limit $\lb\rightarrow\infty$. Solutions exist with $\phi_{i}(0)\sim u$, $\forall~i$, but these require the $Z_2^{(i)}$ symmetry breaking terms to be as ``large'' as the $Z_2^{(i)}$ preserving terms. Calculations become somewhat more cumbersome in this range of parameter space.

For more general model building purposes one may wish to break a discrete symmetry amongst multiple throats to a discrete subgroup; that is one may desire some but not all throat lengths to be equal after stabilization. In the limit $\alpha=\beta$ the present example provides a demonstration of precisely this scenario. Observe from (\ref{throat_length_z3_broke}) that for $\alpha=\beta$ one has $L_2=L_3$. Thus instead of breaking $Z_3$ completely the breaking pattern $Z_3\rightarrow Z_2$ would result. The equality of $\alpha$ and $\beta$ can be motivated by symmetry as in this limit the $Z_3\times Z_2^D$  symmetry of the UV potential is enhanced  to $S_3\times Z_2^D$. Thus the present example can be employed to either break $Z_3$ completely or to break it partially to a $Z_2$ subgroup depending on the relation between $\alpha$ and $\beta$.
%%%%%%%%%%%%%%%%%%%%%%%%%%%%%%%%%%%%%%%%%%%%%%%%%%%%
\section{More Than Three Throats\label{sec:more_than_three}}
The generalization of some of the preceding results to $n>2$ throats is straight forward. For $n$ independent throats one considers $n$ GW scalars with bulk and IR actions matching those of the two independent throat analysis in Section~\ref{sec:2ind_throats}. The common UV action is generalized to
\bea
S_{UV}&=& -\frac{1}{2}\int d^4x \sqrt{-g_{uv}}\left\{\sum_{i=1}^n\bl_i(\phi_i^2-u_i^2)^2+\sum_{i\ne j}\kappa_{ij}\phi_i^2\phi_j^2\right\}.
\eea
In the limit where the $\bl_i$ terms dominate the BCs the resulting potential for the throat lengths $V(\{ L_i\})$ is the sum of $n$ decoupled pieces, each of which  match the GW result to leading order. Thus the leading order expression for the throat lengths that minimize $V(\{ L_i\})$ is:
\begin{eqnarray}
L_i=\frac{1}{\epsilon_i k}\ln\left(\frac{u_i}{v_i}\right).
\eea
As in the two throat case these expressions receive $\mathcal{O}(\kappa_{ij}/\bl)$ corrections as a result of the UV localized interactions amongst scalars.

For $n$ throats related by a $Z_n$ symmetry the results of Sections~\ref{preserving_z2} and~\ref{preserviing_z3} can also be generalized. In those sections, two- and three-throat
systems that preserve $Z_2$ and $Z_3$ throat interchange symmetries were presented. This is generalized by employing bulk scalars in each throat with bulk and IR actions  matching those in Section~\ref{preserviing_z3}. The $Z_n$ symmetric UV action is generalized to\footnote{For a discussion of a similar potential in a different context see~\cite{Foot:2006ru}.}
 \bea
S_{UV}&=& -\frac{1}{2}\int d^4x \sqrt{-g_{uv}}\left\{-\mu^2 \sum_{i=1}^n\phi_i^2+\bl\left[\sum_{i=1}^n\phi_i^2\right]^2+\tl \sum_{i=1}^n\phi_i^4\right\}.
\eea
 For large $\bl,\ \tl,\ \mu>0$ the leading order UV BCs are
\bea
\phi_i^2(0)=\frac{\mu^2}{2(n\bl +\tl)}\equiv u^2,\quad\quad i=1,2,\ .\
.\ ,n.
\eea
and the throat lengths are fixed at
\begin{eqnarray}
L_i=\frac{1}{\epsilon k}\ln\left(\frac{u}{v}\right)\quad,\quad i=1,2,\
.\
.\ ,n,
\eea
thereby preserving the $Z_n$ symmetry and generalizing the earlier results. Just as the $n=3$ case permits a geometric realization of the trinification model the symmetric $n=4$ case would similarly admit a geometric realization of the quartification group~\cite{Joshi:1991yn}. Following the methodology we have presented in the previous sections symmetry breaking scenarios for $n>3$ throats can also be obtained.
%%%%%%%%%%%%%%%%%%%%%%%%%%%%%%%%%%%%%%%%%%%%%%%%%%
\section{Conclusion\label{sec:conc}}
In this work we have generalized the GW mechanism for stabilizing the
single warped throat of the RS model to multithroat backgrounds in
which distinct warped throats share a common UV brane. We have shown
that, due to a combination of IR and UV dynamics, the throat lengths
can be stabilized at finite values in such setups. We considered
independent throats for which the GW results generalize in a straight
forward way and throats related by a discrete interchange symmetry. In
the latter case we provided examples where the stabilization dynamics
can either preserve or break the interchange symmetry. Our results are
applicable to a broad class of multithroat models and have direct
relevance for existing models in the literature. 
%%%%%%%%%%%%%%%%%%%%%%%%%%%%%%%%%%%%%%%%%%%%%%%%%%
\section*{Acknowledgments}
The authors thank D.~P.~George, S.~Gopalakrishna and R.~Volkas for
helpful discussions and D.~Morrissey for comments on the manuscript. They also thank the theory groups at the Nationaal
instituut voor subatomaire fysica (NIKHEF) and the University of Melbourne for
hospitality while parts of this work were completed. SL was
supported in part by the National Science Council of Taiwan. KM was supported
by the National Science and Engineering Research Council of Canada.
%%%%%%%%%%%%%%%%%%%%%%%%%%%%%%%%%%%%%%%%%%%%%%%%%%
%\appendix
%%%%%%%%%%%%%%%%%%%%%%%%%%%%%%%%%%%%%%%%%%%%%%%%%%%

\end{document}